\documentclass[a4paper,11pt]{article}
\pdfoutput=1 

\usepackage{jinstpub} 
\usepackage{amsmath}
\title{\boldmath Improved-RPC for the CMS muon system upgrade for the HL-LHC}

\author[l,1]{Priyanka Kumari,\note{Corresponding author.}}
\author[s]{,K.S. Lee}
\author[n]{,A. Gelmi}
\author[j]{,K. Shchablo}
\author[a]{A. Samalan}
\author[a]{,M. Tytgat}
\author[a]{,N. Zaganidis}
\author[b]{,G.A. Alves}
\author[b]{,F. Marujo}
\author[c]{,F. Torres Da Silva De Araujo}
% ...

\affiliation[a]{Ghent University, Dept. of Physics and Astronomy, Proeftuinstraat 86, B-9000 Ghent, Belgium.}
\affiliation[b]{Centro Brasileiro Pesquisas Fisicas, R. Dr. Xavier Sigaud, 150 - Urca, Rio de Janeiro - RJ, 22290-180, Brazil.}
\affiliation[c]{Dep. de Fisica Nuclear e Altas Energias, Instituto de Fisica, Universidade do Estado do Rio de Janeiro, Rua Sao Francisco Xavier, 524, BR - Rio de Janeiro 20559-900, RJ, Brazil.}
% ...
\affiliation[j]{Univ Lyon, Univ Claude Bernard Lyon 1, CNRS/IN2P3, IP2I Lyon, UMR 5822, F-6922, Villeurbanne, France.}
% ...
\affiliation[l]{Department of Physics, Panjab University, Chandigarh 160 014, India.}
% ...
\affiliation[n]{INFN, Sezione di Bari, Via Orabona 4, IT-70126 Bari, Italy.}
% ...
\affiliation[s]{Korea University, Department of Physics, 145 Anam-ro, Seongbuk-gu, Seoul 02841, Republic of Korea.}
% ...

\emailAdd{priyanka.kumari@cern.ch}

\abstract{
During Phase-2 of the LHC, known as the High Luminosity LHC (HL-LHC), the accelerator will increase its instantaneous luminosity to 5 $\times$ 10$^{34}$ cm$^{-2}$ s$^{-1}$, delivering an integrated luminosity of 3000 fb$^{-1}$ over 10 years of operation starting from 2027. In view of the HL-LHC, the CMS muon system will be upgraded to sustain efficient muon triggering and reconstruction performance. Resistive Plate Chambers (RPCs) serve as dedicated detectors for muon triggering due to their excellent timing resolution, and will extend the acceptance up to pseudorapidity values of $|\eta|$=2.4. Before Long Shutdown 3 (LS3), the RE3/1 and RE4/1 stations of the endcap will be equipped with new improved Resistive Plate Chambers (iRPCs) having different design and geometry than the present RPC system. The iRPC geometry configuration improves the detector's rate capability and its ability to survive the harsh background conditions of the HL-LHC. Also, new electronics with excellent timing performances (time resolution of less than 150 ps) are developed to read out the RPC detectors from both sides of the strips to allow for good spatial resolution along them. The performance of the iRPC has been studied with gamma radiation at the Gamma Irradiation Facility (GIF++) at CERN. Ongoing longevity studies will help to certify the iRPCs for the HL-LHC running period. The main detector parameters such as the current, rate and resistivity are regularly monitored as a function of the integrated charge. Preliminary results of the detector performance will be presented.}


\proceeding{XV Workshop on Resistive Plate Chambers and related detector (RPC2020)\\
  10-14 February 2020\\
  Rome, Italy}

\begin{document}
\maketitle
\flushbottom

\section{Resistive Plate Chambers at CMS}
\label{sec:intro}

The muon system of the Compact Muon Solenoid (CMS) experiment~\cite{CMS} at the CERN Large Hadron Collider (LHC) consists of three types of gaseous detectors: Drift Tubes (DTs) in the barrel, Cathode Strip Chambers (CSCs) in the endcaps and Resistive Plate Chambers (RPCs) in both barrel and endcap regions. A total of 1056 RPC detectors are present in the CMS muon system covering the absolute pseudorapidity region up to $|\eta|$ = 1.9. RPCs~\cite{RPC} are parallel plate gaseous detectors made up of two gas gaps, with a strip readout plane in between, providing an extra coordinate in transverse plane with respect to the beam. The detector operational principle is based on an avalanche mode.

\begin{figure}[htbp]
 \centering
\includegraphics[width=0.8\textwidth]{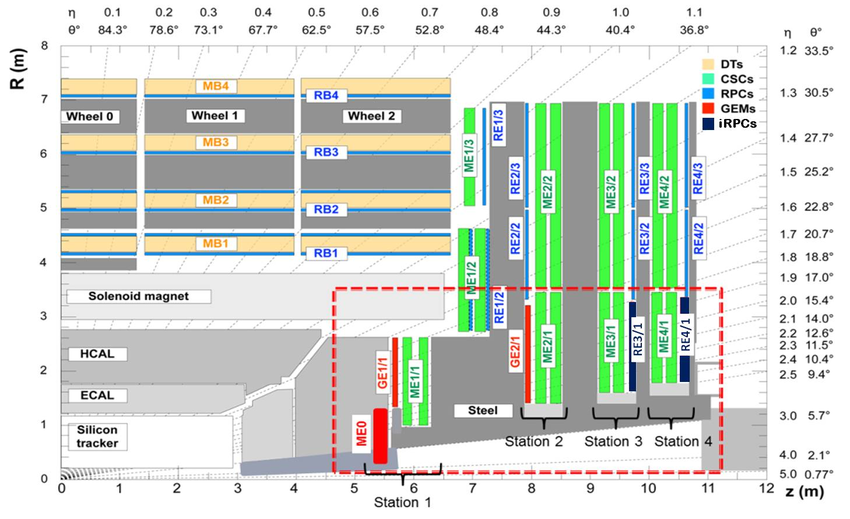}
\vspace{2mm}
\caption{A quadrant of the CMS Muon Spectrometer, showing DT chambers (yellow), RPCs (light blue), and CSCs (green). The locations of new forward muon detectors for the HL-LHC project are contained within the dashed box and indicated in red for Gas Electron Multiplier (GEM) stations (ME0, GE1/1, and GE2/1) and violet for improved RPC stations (RE3/1 and RE4/1).}
\label{fig:1}
\end{figure}

The RPCs have an excellent time resolution on the order of 1-2 ns, making them suitable for muon triggers~\cite{timeres}. The gas mixture used in the operation of the RPCs is 95.2\% C${_2}$H${_2}$F${_4}$ (tetrafluoroethane), 4.5\% i-C${_4}$H${_{10}}$ (isobutane), and 0.3\% SF${_6}$ (sulphur hexafluoride).

To ensure good redundancy and excellent triggering during the second phase of the LHC physics program, known as the High Luminosity LHC (HL-LHC), the RPC system needs to be upgraded~\cite{HL-LHC}. During the HL-LHC, instantaneous luminosity will increase 5 times compared to the 
present nominal luminosity, which will result in a harsher background rate. The RPC muon upgrade project will increase redundancy and robustness by installing 72 improved RPC (iRPC) in stations 3 and 4 in the pseudorapidity region 1.9 $< |\eta| <$ 2.4 as shown in Figure~\ref{fig:1}. 

\section{Motivation behind the improved RPC}
\begin{figure}[h!]
  \begin{center}
    \includegraphics[height=5.5cm,width=7cm]{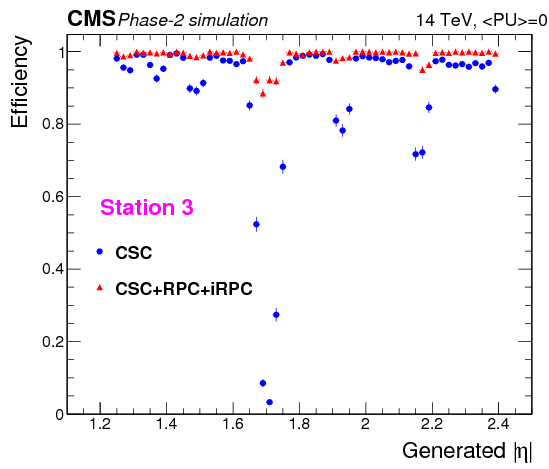}
    \qquad
    \includegraphics[height=5.5cm,width=7cm]{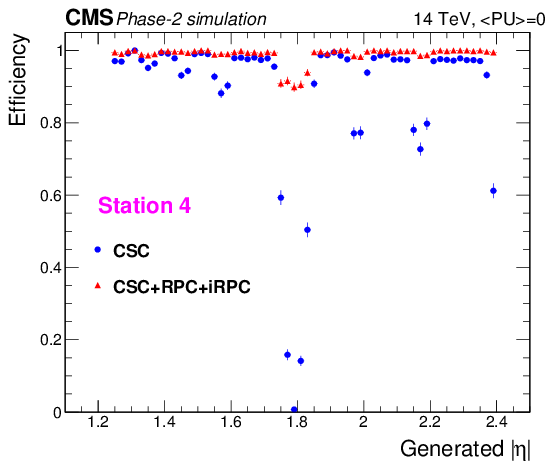}
    \caption{Comparison between L1 single muon trigger primitive efficiencies with and without RPC hits as a function of $|\eta|$ for stations 3 (left) and 4 (right). The contribution of iRPC starts after $|\eta|$ = 1.9.}
    \label{fig:2}
  \end{center}
\end{figure}

Despite having CSCs in ME3/1 and ME4/1, there is a sharp drop in muon trigger efficiency at few $|\eta|$ regions but adding the RE3/1 and RE4/1 stations along with CSCs will enhance the local muon measurement by adding track hits and by increasing the lever arm. In addition, they will mitigate ambiguities in resolving the multiple tracks in the endcap trigger. Furthermore, better background rejection and capability of reconstruction for slowly moving heavy stable charged particles will be achieved by adding the RE3/1 and RE4/1 stations to the CSC detectors. Figure~\ref{fig:2} shows the trigger primitive efficiencies in the RE3/1 (left) and the RE4/1 (right) regions with (red) and without (blue) the presence of RPC trigger information. An improvement of the trigger efficiency at the level of 15\% is expected when the RPC hits are added to the L1 single muon trigger.

\section{Design and Specifications of new improved RPC chambers}
In the HL-LHC phase, the average background rate expected in the RE3 and RE4 stations is about 600 Hz/cm$^{2}$. Considering a safety factor of 3, the required rate capability of the iRPCs will be $\approx$ 2 kHz/cm$^{2}$. To cope with high background rates in the HL-LHC, we have pursued a higher detector sensitivity compared to the RPCs presently installed in CMS. In past years, many prototype detectors based on the current phenolic RPC technology have been constructed with different thicknesses of gas gaps and RPC electrodes and tested at the KOrean DEtector Laboratory (KODEL)~\cite{iRPC}. The rate capability of the RPCs can be improved by reducing the recovery time of the electrodes and the total charge produced in a discharge.

\begin{figure}[htbp]
  \centering
  \includegraphics[width=0.48\textwidth]{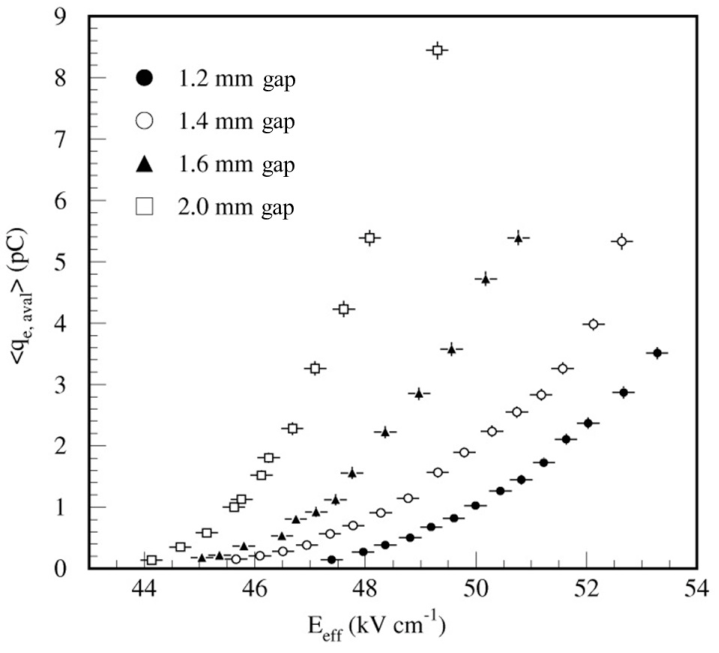}
  \vspace{2mm}
  \caption{Average charge per avalanche measured on 1.2 mm (full circles), 1.4 mm (open circles), 1.6 mm (triangles), and 2.0 mm (squares) double-gap RPCs, as a function of the electric field strength.}
  \label{fig:3}
\end{figure}

Figure~\ref{fig:3}~\cite{charge} shows the average pick-up charge drawn in double-gap RPCs with 4 different gas gap thicknesses, ranging between 1.2 and 2.0 mm, as a function of the electric field strength. With the thinner gas gap, the path gets reduced and thus following the Townsend formula reduces exponentially the gain. However increasing slightly the electric field by increasing the applied voltage allows one to reach an adequate gain and enough charge within the avalanche. The operational condition of the RPCs with the thinner thicknesses for the gap and the electrodes is essential to suppress the fast increase of cluster sizes with increasing electric field because the operational HV is realistically limited by the cluster sizes. In addition, the range of the lateral spread of the electromagnetic induction is narrower with thinner thickness of the electrodes and of the gas gap.
The operational plateau can be defined by the range where the efficiency is higher than 95\%. The choice of the thinner thickness of gaps and electrodes provides us an essential condition when we try to suppress fast increase of cluster size measured with narrow pitch strips (less than 10 mm) at lower digitization thresholds. The reduction of the operational high voltage will reduce the risk of detector aging and also improve the robustness of the high voltage system. Also the induced charge depends on the ratio between the gas gap and the electrode thickness. The thicker the electrode the lower the pick-up charge. So another good reason to lower the electrode thickness, helping the Front-End electronics. For the iRPCs, a gap thickness of 1.4 mm is chosen as safe compromise, taking into account that thinner gaps would be more sensitive to non-uniformities.
\begin{figure}[htbp]
  \centering
  \includegraphics[width=0.7\textwidth]{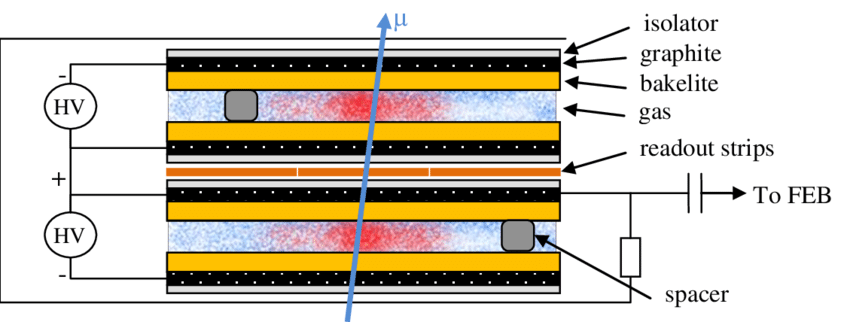}
  \vspace{2mm}
  \caption{Double gap RPC design.}
  \label{fig:4}
\end{figure}

As shown in Figure~\ref{fig:4}, an iRPC is composed of two gas gaps sandwiching a pick-up strip plane in the middle. The thickness of the gas gap and the RPC electrode (Bakelite) is 1.4 mm. The High Pressure Laminate (HPL) resistivity has been specified to be in the range from 0.9 to 3 $\times$ 10$^{10}~ \Omega$-cm, a factor of 2 less than that of the present RPCs system.

In order to improve the detector sensitivity and the rate capability of the iRPCs, we have chosen a thinner thickness for the gaps and the electrodes, which effectively suppress the fast increase of cluster size with the applied HV when the detector is operated with the higher sensitive front-end-electronics~\cite{petiroc}. The iRPCs are equipped with a charge sensitive Front-End-Electronics board (FEB) whose charge sensitivity is at least 5 times better than the ones used for the present RPCs. The currently proposed value of the digitization threshold to be applied to the iRPC pulses is 50 fC. However, we plan to decrease the charge sensitivity of the FEB to values as low as 20 fC in the future.

\subsection{iRPC front-end electronics and its validation at GIF++}
\begin{figure}[htbp]
\centering
  \includegraphics[width=0.32\textwidth]{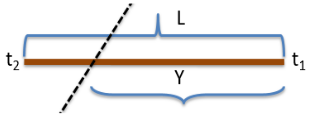}
  \vspace{2mm}
  \caption{Diagram to determine the hit position using time information from both ends of the strip.}
  \label{fig:5}
\end{figure}

The new FEB for the iRPC is equipped with an ASIC PETIROC and a field-programmable gate-array (FPGA) Altera Cyclone II. The ASIC chip, based on SiGe technology, contains 32 fast channel preamplifiers operating with a gain of 25 and with an overall bandwidth of 1 GHz. The FPGA Cyclone II includes a time-to-digital-converter (TDC) to measure the signal travel time. The signals are read from both strip ends. The time difference between the signals coming from the two ends of the strip ({\it t$_{2}$-t$_{1}$}) is used to determine the hit position ({\it Y)} of the particle along the strip ($\eta$ position) as given in equation~\ref{equ:1} and shown in Figure~\ref{fig:5}.
\begin{equation}
  Y = L/2 - v \times (t_{2} - t_{1})/2
  \label{equ:1}
\end{equation}
where {\it t$_{2}$} and {\it t$_{1}$} are the arrival times from both ends of the strip, {\it v} is the signal propagation velocity, and {\it L} is the strip length.\\
\begin{figure}
  \centering
  \includegraphics[width=0.47\textwidth]{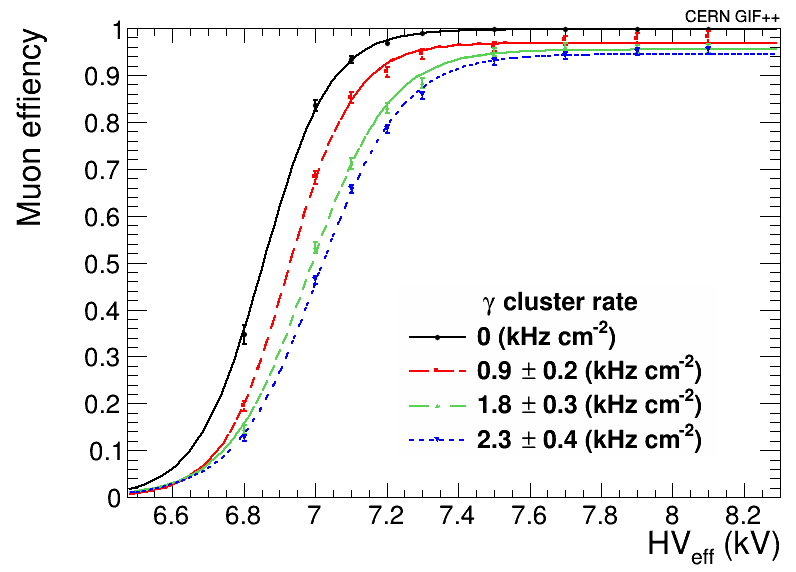}
  \includegraphics[width=0.49\textwidth]{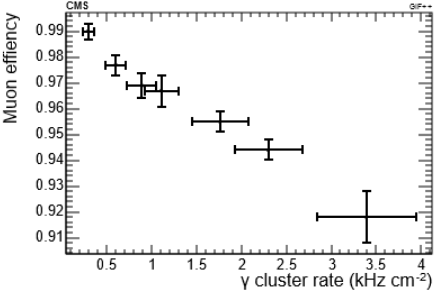}
  \vspace{2mm}
  \caption{Left: Efficiencies as a function of the effective high voltage measured at different background rates. Right: Evolution of the efficiency at working point at different background cluster rates. The measurements were performed with a threshold of 81 fC.}
  \label{fig:6}
  \end{figure}
A full-size iRPC prototype with new FEB and readout Printed Circuit Board (PCB) having average strip pitch of 0.75 cm was tested at the Gamma Irradiation Facility (GIF++)~\cite{gif} at CERN with a muon beam under varying radiation levels. To examine the rate capability of a full-size prototype iRPC, we exploited the current activity of the $^{137}$Cs gamma source which is 13.9 TBq. The iRPC prototype equipped with the new FEBs was tested with gamma rates higher than 2 kHz/cm$^{2}$. At this background rate, a factor 3 higher than that expected in the HL-LHC, the efficiencies obtained for cosmic muons were measured to be reliably higher than 95\%.

The left plot in Figure~\ref{fig:6} shows the efficiency as a function of effective high voltage measured at 4 different background rates and the right plot shows the working point efficiency as a function of the background cluster rate. The working point is defined by standard definition of RPC.
\begin{equation}
  \epsilon(HV_{eff}) = \frac{\epsilon_{max}}{1 + e^{-\lambda(HV_{eff} - HV_{50})}}
\end{equation}
where $\epsilon_{max}$ is the maximal efficiency of the detector, $\lambda$ is proportional to the slope at half maximum and $HV_{50}$ is the voltage value at which the efficiecny reaches half of the maximum.

The sigmoid function allows to define the two important parameters: the ``knee'' defined as the voltage at 95\% of the maximum efficiency, and the ``working point'' defined as the:
\begin{equation}
  WP = HV_{knee} + 150~V
\end{equation}
The WP is shifted by $\approx {200}$ V for each background rate between 0-2 kHz.


\subsection{iRPC sensitivity and background rate study at the HL-LHC}
\begin{figure}
  \centering
  \includegraphics[width=0.4\textwidth]{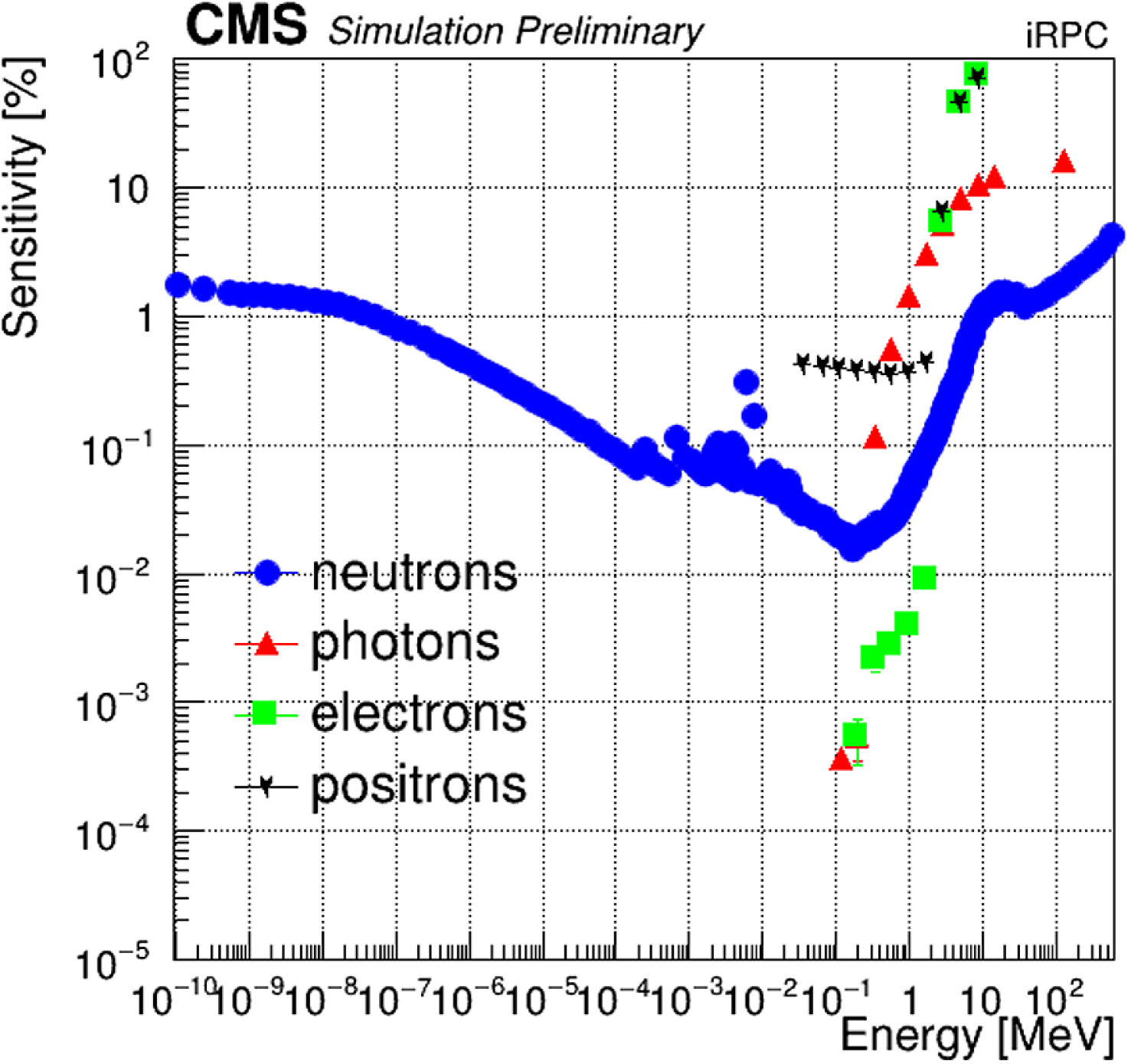}
  \vspace{2mm}
  \caption{iRPC sensitivity as a function of the kinetic energy of the incident particles.}
  \label{fig:7}
\end{figure}

\begin{figure}
  \centering
  \includegraphics[width=0.49\textwidth]{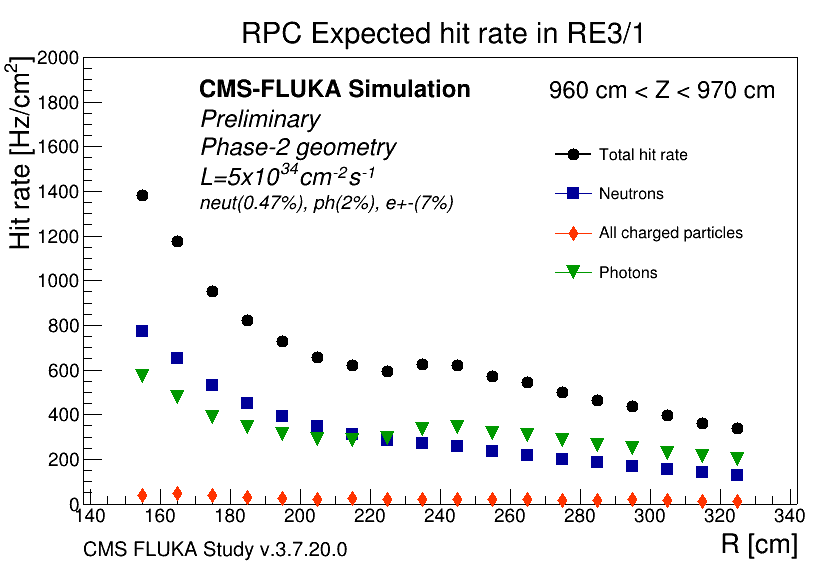}
  \includegraphics[width=0.49\textwidth]{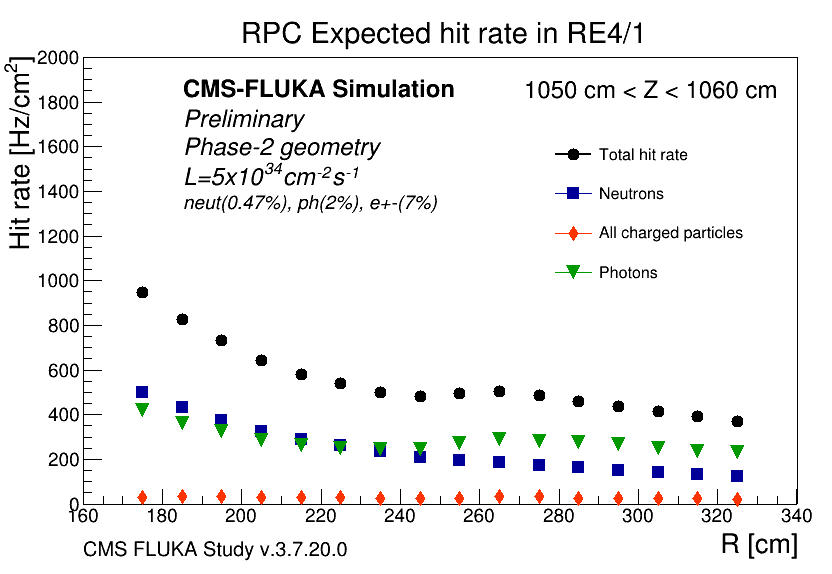}
  \vspace{2mm}
  \caption{RPC expected hit rates in RE3/1 (left) and RE4/1 (right) during the HL-LHC as a function of the distance (R) from the center of the CMS beam pipe.}
  \label{fig:8}
\end{figure}

The iRPC geometry is new, so to estimate the expected background rate it is necessary to study the iRPC sensitivity. The GEANT4 simulation toolkit~\cite{geant} was used to simulate the iRPC geometry and then to study its sensitivity. Sensitivity ({\it S}) indicates how the detector responds to the background particles ({\it N}$_{BG}$) and is defined as the probability for a particle, at a given energy reaching the detector surface, to produce a signal ({\it N}$_{Hit}$) as given in Equation~\ref{equ:2}.
\begin{equation}
  S(E) = \frac{N_{Hit}}{N_{BG}}(E)
  \label{equ:2}
\end{equation}
The sensitivity is a function of the energy of the incident particles because at different energies, different processes are responsible for the production of secondary particles~\cite{sensitivity}. The iRPC sensitivity has been studied with different particles that constitute the CMS background at the energy expected during the HL-LHC as shown in Figure~\ref{fig:7}.
Sensitivity to the neutrons ({\it n}) increases at low energy because of the well known fact that the cross section of gammas coming from the ({\it n}, $\gamma$) capture reaction increases with decreasing neutron energy as $\sigma \propto $ 1/$\sqrt{E}$.\\
The expected background hit rates in the RE3/1 and RE4/1 stations during the HL-LHC have been estimated by the incident particle fluxes along with iRPC sensitivity. The incident energy fluxes for various background particles have been estimated using FLUKA simulations~\cite{fluka}, which are used to describe the upgraded CMS geometry and to propagate particles through the detector and surrounding material~\cite{hitrate}. The expected background hit rate at an instantaneous luminosity of 5 $\times$ 10$^{34}$ cm$^{-2}$s$^{-1}$ can be plotted as a function of the distance (R) from the center of the CMS beam pipe for both the RE3/1 and RE4/1 endcap stations, as shown in Figure~\ref{fig:8}.
Finally, the average background rate has been estimated using the average sensitivity values, [neutron (0.4\%), $\gamma$ (2.5\%), e$^{+\textbackslash-}$ (6.8\%)]. The average background rate of $\approx$ 600 Hz/cm$^{2}$ defines the requirement of a minimum rate capability of $\approx$ 2 kHz/cm$^{2}$ for the iRPCs.

\section{Summary and Conclusion}
In the HL-LHC, the background rates in the RE3/1 and RE4/1 regions are expected to be 10 times higher than those we have experienced in the existing endcap RPCs. To maintain the robustness and redundancy as well as the identification and reconstruction capabilities of muon system, the RPC upgrade project is presented. The 72 new improved RPCs will be installed in the innermost rings of stations 3 and 4, corresponding to the pseudorapidity range of 1.9 $< |\eta| <$ 2.4 during the technical stops before LS3.

The thickness of the gas gap and of the RPC electrodes is chosen as 1.4 mm rather than 2 mm, which has been used for the existing RPC system. The choice of the thinner thickness is to ensure a shorter removal time of the avalanche charge through the RPC electrodes which enhances the rate capability. Also, use of a lower threshold permits to preserve the size of the operational plateau of the previous RPC version. The reduction of operational high voltage will reduce the risk of detector aging and helps to improve robustness of the high voltage system.\\
The present iRPC prototype module equipped with new front-end electronics has been examined using intensive gamma background at GIF++. It fulfills the requirements of operation in the high background rates of the HL-LHC phase: the efficiency at particle rates of 2 kHz/cm$^{2}$ which is (a factor of 3 higher than the mean value expected in the iRPCs in future HL-LHC runs) is higher than 95\%.

\acknowledgments

We would like to acknowledge the Council of Scientific and Industrial Research (CSIR), India and the EHEP group of the Department of Physics, Panjab University Chandigarh for providing the funds for the main analyst to attend this conference. This project has received funding from the European Union's Horizon 2020 Research and Innovation programme under Grant Agreement no. 654168. Also sincere thanks to the RPC2020 organizers and the RPC community for a very successful conference.


\end{document}